\documentclass[fleqn,twoside]{article}
\usepackage{espcrc2}

\newcommand{\be}{\begin{equation}}
\newcommand{\ee}{\end{equation}}
\newcommand{\bea}{\begin{eqnarray}}
\newcommand{\eea}{\end{eqnarray}}
\newcommand{\nn}{\nonumber \\}
\newcommand{\half}{\frac{1}{2}}

\newcommand{\AmS}{{\protect\the\textfont2
  A\kern-.1667em\lower.5ex\hbox{M}\kern-.125emS}}

\title{Supersymmetry and Gravity in Noncommutative Field Theories}

\author{{Victor O. Rivelles \\Center for Theoretical Physics \\
  Massachusetts Institute of 
  Technology \\ Cambridge, MA  02139, USA}\thanks{On leave from
  Instituto de F\'\i sica, Universidade de
  S\~{a}o Paulo, Caixa Postal 66318, 05315-970, S\~{a}o Paulo, SP,
  Brazil; e-mail: rivelles@lns.mit.edu}}
       
\begin{document}

\begin{abstract}
We discuss the renormalization properties of noncommutative
supersymmetric theories. We also discuss how the gauge field plays a 
role similar to gravity in noncommutative theories.
\end{abstract}

\maketitle

\section{Introduction}

Non commuting coordinates have been proposed long ago as a way to get
rid of divergences in quantum field theory \cite {Jackiw:2003dw} and
more recently in the context of string theory with D-branes
\cite{Seiberg:1999vs}. Its effects can easily be seen. If $x$ and $y$
have a commutation relation $[x,y]=\theta$ then at the quantum level
we expect an uncertainty $\Delta x \, \Delta y \sim \theta$. Together
with the usual uncertainty relation $\Delta x \, \Delta p_x \sim 1$ we
find that $\Delta y \sim \theta \Delta p_x$. This means that the
ultraviolet (UV) regime in the $x$-direction produces an infrared (IR)
effect in the $y$-direction and vice-versa. At the quantum field
theory level this phenomenon manifests itself as a mixture of UV
and IR divergences already at the one loop level
\cite{Minwalla:1999px}. If the theory is renormalized in the usual way
IR divergences appear jeopardizing renormalizability at higher loop
levels. It was suggested that supersymmetry would improve this
situation since they are less divergent than conventional theories
\cite{Ruiz:2000hu} and this was found to be true for the gauge field
two-point function at one loop level \cite{Matusis:2000jf}. A general
proof soon appeared showing that the noncommutative (NC) Wess-Zumino
model is free of the UV/IR mixing to all orders in perturbation
theory \cite{Girotti:2000gc}. So far, it is the only known model of a
four dimensional renormalizable noncommutative (NC) field theory. Its
low energy properties were studied in detail \cite{Girotti:2001dh}. Other 
noncommutative supersymmetric non-gauge theories were also found to be
free of UV/IR mixing. For instance, the supersymmetric nonlinear sigma
model in three dimensions turns out to be renormalizable in the
$1/N$ expansion \cite{Girotti:2001gs,Girotti:2002kr}. Spontaneous
symmetry breaking also has troubles in the presence of
noncommutativity \cite{Campbell:2000ug}. In three dimensions the
situation is improved and it seems that supersymmetry plays no role in
this case \cite{Girotti:2002kr}. For supersymmetric gauge theories up
to two loop orders the mixing is also absent \cite{Girotti:2002se} but
it seems that for higher loops this is no longer true. 

Another interesting aspect of NC field theories is the Seiberg-Witten
(SW) map. Instead of working with NC fields with its exotic properties we
map them to ordinary commutative fields \cite{Seiberg:1999vs}. In this
way a local field theory is obtained at the expense of introducing
non-renormalizable interactions. Many properties of the NC fields are
usually lost after performing the SW map. For instance, translations
in NC directions are 
equivalent to gauge transformations \cite{Gross:2000ph}, a feature 
similar to that found in general relativity where local translations
are associated to general coordinate transformations. This property is
completely lost after the SW map. However, we found that another
aspect concerning gravity emerges: NC theories can be interpreted as
ordinary theories immersed in a gravitational background generated by
the gauge field \cite{Rivelles:2002ez}. What is interesting is that
the gravity coupling is sensitive to the charge with
uncharged fields coupling more strongly than charged ones. 

In the next section we will review the UV/IR mixing for the simple
case of a scalar field showing how the IR divergences become a source
of trouble. Then, in section 3 we will discuss the NC
Wess-Zumino model proving its renormalizability to all loop
orders. Section 4 is devoted to the study of the UV/IR mixing in the
NC Gross-Neveu and nonlinear sigma models. In section 5 we show how
supersymmetry improves the situation in the NC supersymmetric
nonlinear sigma model. Aspects concerning spontaneous symmetry
breaking in NC theories are discussed in section 6 and the relation
between gravitation and NC theories is discussed in the last section. 

\section{UV/IR Mixing for the Scalar Field}

Noncommutative field theories are obtained from the commutative ones
by replacing the ordinary field multiplication for the Moyal product,
which is defined as
\be
\label{4}
\left( \phi_1 \star \phi_2 \right) (x) \equiv \left[ e^{\frac{i}{2}
    \theta^{\mu\nu} \frac{\partial}{\partial x^\mu}
    \frac{\partial}{\partial y^\nu} } \phi_1(x) \phi_2(y)
\right]_{y=x}.
\ee
Then the noncommutative $\phi^4$ model in $3+1$ dimensions reads as
    \cite{Minwalla:1999px}   
\be
L =  \half \partial_\mu \phi \star
    \partial^\mu \phi - \frac{m^2}{2} \phi \star \phi 
- \frac{g^2}{4!}
\phi\star\phi\star\phi\star\phi.
\ee

We now proceed in the standard way. The quadratic terms gives the
propagator. Since the Moyal product has the property 
$
\int dx \,\, (f \star g)(x)  =
\int dx \,\, f(x) g(x),
$
the propagator is the same as in the commutative case. This is a
general property of noncommutative theories: the propagators are not 
modified by the noncommutativity. The vertices, however, are in
general affected by phase factors. In this case we get 
 \bea
&& - \frac{g^2}{6}[\cos(\half k_1 \wedge k_2) \cos (\half k_3 \wedge
   k_4) + \nn   
&&\cos(\half k_1 \wedge k_3) \cos (\half k_2 \wedge k_4) + \nn
&& \cos(\half k_1 \wedge k_4) \cos (\half k_2 \wedge k_3) ]. 
\eea

We can now compute the one loop correction for the two-point
function. It is easily found to be 
\be 
\frac{g^2}{3 (2 \pi)^4} \int d^4k \,\, \left( 1 + \half \cos(k \wedge
  p) \right) \frac{1}{k^2 + m^2}.
\ee
The first term is the usual one loop mass correction of the
commutative theory (up to a factor $1/2$) which is quadratically
divergent. The second term is not divergent due to the oscillatory
nature of $\cos(k \wedge 
p)$. This shows that the non-locality introduced by the Moyal product
is not so bad and leaves us with the same divergence structure of the
commutative theory. This is also a general property of noncommutative
theories \cite{Filk}. To take into account the effect
of the second term we regularize the integral using the Schwinger
parametrization 
\be
\frac{1}{k^2 +m^2} = \int_0^\infty d\alpha \,\, e^{- \alpha (k^2 +
  m^2)} e^{-\frac{1}{\Lambda^2 \alpha}},
\ee
where a cutoff $\Lambda$ was introduced. We find
\bea
&& \Gamma^{(2)} = \frac{g^2}{48 \pi^2} [ ( \Lambda^2 - m^2
\ln(\frac{\Lambda^2}{m^2}) + \dots ) + \nn
&&\half ( \Lambda^2_{eff} - m^2
\ln(\frac{\Lambda^2_{eff}}{m^2}) + \dots ) ],
\eea
where
$
\Lambda^2_{eff} = 1/(\tilde{p}^2 + 1/\Lambda^2), 
\tilde{p}^\mu = \theta^{\mu\nu} p_\nu.
$
Note that when the cutoff is removed, $\Lambda \rightarrow \infty$,
the noncommutative contribution remains finite providing a natural
regularization. Also $\Lambda^2_{eff} = 1/\tilde{p}^2$ which
diverges either when  $\theta \rightarrow 0$ or when $\tilde{p}
\rightarrow 0$. 

The one loop effective action is then
\bea
&&\int d^4p \,\, \half ( p^2 + M^2 + \frac{g^2}{96 \pi^2
(\tilde{p}^2 + 1/\Lambda^2)} - \nn 
&&\frac{g^2 M^2}{96 \pi^2}
\ln\left(\frac{1}{M^2(\tilde{p}^2 + 1/\Lambda^2)} \right))
  \phi(p) \phi(-p),
\eea
where $M$ is the renormalized mass. Let us take the limits $\Lambda
\rightarrow \infty$ and $\tilde{p} \rightarrow 0$. If we take first
$\tilde{p} \rightarrow 0$ then $\tilde{p}^2 << 1/\Lambda^2$
and $\Lambda_{eff}=\Lambda$ showing that we 
recover the effective commutative theory. However, if we take $\Lambda
\rightarrow \infty$ then $\tilde{p}^2 >> 
1/\Lambda^2$ and $\Lambda^2_{eff} = 1/\tilde{p}^2$
so that we get  
\bea 
&&\int d^4p \half ( p^2 + M^2 + \frac{g^2}{96 \pi^2 \tilde{p}^2}
- \nn
&&  \frac{g^2 M^2}{96 \pi^2} \ln\left(\frac{1}{M^2 \tilde{p}^2}\right) +
  \dots ) \phi(p) \phi(-p),
\eea
which is singular when $\tilde{p} \rightarrow 0$. 
This shows that the limit $\Lambda \rightarrow \infty$ does not
commute with the low momentum limit $\tilde{p} \rightarrow 0$  so that
there is a mixing of UV and IR limits. 

The theory is renormalizable at one loop order if we do not
take $\tilde{p} \rightarrow 0$. What about higher loop orders? Suppose
we have insertions of one loop mass corrections. Eventually we will have
to integrate over small values of $\tilde{p}$ which diverges when
$\Lambda \rightarrow \infty$. Then we find an IR
divergence in a massive theory. This combination of UV and IR
divergences makes the theory non-renormalizable. 


Then the main question now is the existence of a theory
which is renormalizable to all loop orders. Since the UV/IR mixing
appears at the level of quadratic divergences a candidate theory would
be a supersymmetric one because it does not have such divergences. As
we shall see this indeed happens.  

\section{Noncommutative Wess-Zumino Model}

The noncommutative Wess-Zumino model in $3+1$ dimensions
\cite{Girotti:2000gc} has an interaction Lagrangian given by 
\bea
 {\cal L}_g &=&  g (F\star A \star
 A-F\star B \star B + G\star A \star B + \nn 
 &&G \star B \star A - \overline \psi
 \star \psi\star A - \overline \psi\star  i\gamma_5 \psi \star B) \nonumber,
 \eea
where $A$ and $B$ are bosonic fields, $F$ and $G$ are auxiliary
fields and $\psi$ is a Majorana spinor. The quadratic part of the
 Lagrangian is identical to the commutative case. The action is
 invariant under the usual supersymmetry transformations. The supersymmetry
transformations  are not modified by the 
Moyal product since they are linear in the fields. 

As usual, the propagators are not modified by noncommutativity and 
the vertices are modified by phase factors. 
The degree of superficial divergence for a generic 1PI graph
$\gamma$ is then 
$
d(\gamma)= 4 -  I_{AF} -I_{BF}-N_A-N_B-2 N_F-2N_G - \frac32 N_\psi, 
$
where $N_{\cal O}$ denotes the number of external lines
associated to the  field ${\cal O}$ and $ I_{AF}$ and $I_{BF}$
are the numbers of internal lines associated to the mixed
propagators $AF$ and $BF$, respectively. In all cases we will
regularize the divergent Feynman integrals by assuming that a 
supersymmetric regularization scheme does exist.

The one loop analysis can be done in a straightforward way. 
As in the commutative case all tadpoles contributions add up to
zero. We have verified this explicitly. The self-energy of $A$ can
be computed and the divergent part is contained in the integral
\be
16 g^2\int\frac{d^4k}{(2\pi)^4}( 1 + \half \cos(k \wedge p) ) \frac{(p \cdot
  k)^2}{(k^2- m^2)^3}.
\ee
The first term is logarithmically divergent. It differs by a factor 2
from the commutative case. As usual, this 
divergence is eliminated by a wave function renormalization. 
The second term is UV convergent and for small $p$ it behaves as 
$p^2 \ln (p^2/m^2)$ and actually vanishes for $p=0$. Then there is no
IR pole. The same analysis can be carried out for the others
fields. 
Therefore, there is no UV/IR mixing in the self-energy as
expected. 

To show that the model is renormalizable we must also look into the
interactions vertices. The $A^3$ vertex has no divergent parts as in
the commutative case. The same happens for the other three point
functions. For the four point vertices no divergence is found as in
the commutative case. Hence, the noncommutative Wess-Zumino model is
renormalizable at one loop with a wave-function renormalization and no
UV/IR mixing.  

To go to higher loop orders we proceed as in the commutative case. We
derived the supersymmetry Ward identities  
for the n-point vertex function. Then we showed that there is a
renormalization prescription which is consistent with the Ward
identities. They are the same as in the commutative case. And finally
we fixed the primitively divergent vertex functions. Then we found
that there is only a common wave function renormalization as in the
commutative case. In general we expect 
$
\varphi_R = Z^{-1/2} \varphi, m_R = Z m + \delta m, 
g_R = Z^{3/2} Z^\prime g.
$
At one loop we found { $\delta m = 0$ and $Z^\prime = 1$}. We showed
that this also holds to all orders and no mass renormalization is
needed. 

Being the only consistent noncommutative quantum field theory in $3+1$
dimensions known so far it is natural to study it in more detail. As a
first step in this direction we considered the non-relativistic limit
of the noncommutative Wess-Zumino model \cite{Girotti:2001dh}. We found
the low energy 
effective potential mediating the fermion-fermion and boson-boson
elastic scattering in the non-relativistic regime. Since
noncommutativity breaks Lorentz invariance we formulated the theory in
the center of mass frame of reference where the dynamics simplifies
considerably. For the fermions we found that the potential is
significantly changed by the noncommutativity while no modification
was found for the bosonic sector. The modifications found give rise to
an anisotropic differential cross section. 

Subsequently the model was formulated in superspace and again found to
be renormalizable to all loop orders \cite{Bichl:2000zu}. The one and two
loops contributions to the effective action in superspace were also
found \cite{Buchbinder:2001hn}. The one loop Kahlerian effective potential
does not get modified by noncommutativity and the two loops non-planar
contributions to the Kahlerian effective potential are leading in the
case of small noncommutativity \cite{Buchbinder:2001hn}.

\section{Noncommutative Gross-Neveu and  Nonlinear Sigma Models}

Another model where renormalizability is spoiled by the
noncommutativity is the $O(N)$ Gross-Neveu model. The commutative
model is 
perturbatively renormalizable in $1+1$ dimensions and $1/N$
renormalizable in $1+1$ and $2+1$ dimensions. In both cases it
presents dynamical mass generation. It is described by the Lagrangian 
\begin{equation}
{\cal L}=\frac{i}2\overline \psi_i \not \!\partial \psi_i +\frac
{g}{4N}(\overline \psi_i \psi_i)(\overline \psi_j \psi_j),\label{1}
\end{equation}
where $\psi_i, i=1,\ldots N$, are two-component Majorana
spinors. Since it is renormalizable in the $1/N$ expansion in $1+1$
and $2+1$ dimensions we will consider both cases. As usual, we
introduce an auxiliary field $\sigma$ and the Lagrangian turns into 
\begin{equation}
{\cal L} =\frac{i}2\overline \psi_i \not \!\partial \psi_i -\frac{\sigma}2 
(\overline \psi_i \psi_i)- \frac{N}{4g}\sigma^2.\label{2}
\end{equation}
Replacing $\sigma$ by $\sigma+M$ where $M$ is the VEV of the
original $\sigma$ we get the gap equation (in Euclidean space) 
$
 {M}/{2g}- (1/2\pi)^D\int {d^Dk}{M}/({k^{2}_{E}+M^2})=0.\label{5}
$
To eliminate the UV divergence we need to renormalize the coupling
constant by 
$
{1}/{g}= {1}/{g_R} + (2/{(2\pi)^D}) \int {d^Dk}/({k^{2}_{E}
+\mu^2}). \label{6}
$
In $2+1$ dimensions we find 
$
{1}/{g_R}= ({\mu-|M|)/}{2\pi}, \label{7}
$
and therefore only for $-{1}/{g_R}+{\mu}/{2\pi}>0$ it is possible to
have $M\not =0$, otherwise $M$ is necessarily zero. 
No such a restriction exists in $1+1$ dimensions. In any case, 
we will focus only in the massive phase.
The propagator for $\sigma$ is proportional to the inverse of the
following expression
$$
 -\frac {iN}{2g}-  iN \int \frac{d^Dk}{(2\pi)^D} \frac{k\cdot (k+p) + 
M^2}{(k^2-M^2)[(k+p)^2-M^2]}, \label{8}
$$
which is divergent. Taking into account the gap equation the above
expression reduces to 
$$
\frac{(p^2-4M^2)N}2\int\frac{d^Dk}{(2\pi)^D} \frac{1}{(k^2-M^2)[(k+p)^2-M^2]},
\label{10}
$$
which is finite. Then there is a fine tuning which is responsible for
the elimination of the divergence and which might be absent in the
noncommutative case due to the UV/IR mixing. 

The interacting part of the noncommutative model is defined by
\cite{Girotti:2001gs}  
\be
L_i = \frac12\sigma \star(\overline \psi\star
\psi)-  \frac{N}{4g}\sigma^2- \frac{N}{2g}M\sigma. 
\ee
Elimination of the auxiliary field results in a four-fermion
interaction of the type $\overline \psi_i\star\psi_i\star\overline
\psi_j\star\psi_j$. However a more general four-fermion interaction
may involve a term like $\overline \psi_i\star\overline
\psi_j\star\psi_i\star\psi_j$. This last combination does not have a
simple $1/N$ expansion and we will not consider it. The Moyal product
does not affect the propagators and the trilinear vertex acquires a
correction of $\cos(p_1\wedge p_2)$ with regard to the commutative
case. Hence the gap equation is not modified, while in the propagator for
the $\sigma$ we find a divergent piece.

On the other side, the nonlinear sigma model also presents troubles in
its noncommutative version. The noncommutative model is described by 
\begin{equation}\label{121}
{\cal L} =-\frac12\varphi_i (\partial^2 + M^2)\varphi_i + \frac12
\lambda\star \varphi_i \star \varphi_i - \frac{N}{2g}\lambda,
\end{equation}
where $\varphi_i$, $i=1,\ldots , N$, are real scalar fields, $\lambda$
is the auxiliary field and $M$ is the generated mass. The leading
correction to the $\varphi$ self-energy is
\begin{equation}\label{122}
-i\int \frac{d^2 k}{(2\pi)^2} \frac{\cos^2 (k\wedge
 p)}{(k+p)^2-M^2}\Delta_\lambda(k), 
\end{equation}
where $\Delta_\lambda$ is the propagator for $\lambda$. As for the
case of the scalar field this can be decomposed as a sum of a
quadratically divergent part and a UV finite part. Again there is the
UV/IR mixing destroying the $1/N$ expansion. 

\section{Noncommutative Supersymmetric Nonlinear Sigma Model}

The Lagrangian for the commutative supersymmetric sigma
model is given by 
\bea
&&L = \frac12 \partial^\mu \varphi_i \partial_\mu \varphi_i +
\frac{i}{2} \overline \psi_i \not \! \partial \psi_i + \frac12 F_i F_i
+ \sigma \varphi_i F_i  \nn && 
 + \frac12\lambda 
\varphi_i \varphi_i  - \frac12\sigma \overline \psi_i \psi_i -
\overline \xi \psi_i \varphi_i - \frac{N}{2g}\lambda,
\eea
where $F_i$, $i=1,\ldots, N$, are auxiliary fields. Furthermore,
$\sigma,\lambda$ and $\xi$ are the Lagrange multipliers which
implement 
the supersymmetric constraints. After the change of variables 
$\lambda\rightarrow \lambda + 2 M \sigma$, $F\rightarrow F-M\varphi$
where $M=<\sigma>$, and the shifts $\sigma\rightarrow \sigma +M$ and
$\lambda\rightarrow \lambda + \lambda_0$, where $\lambda_0=<\lambda>$,
we arrive at a more symmetric form for the Lagrangian 
Now supersymmetry requires $\lambda_0=-2M^2$ and the gap equation is 
$
(1/{(2\pi)^D})\int {d^D k} {i}/({k^2-M^2})= {1}/{g},\label{16}
$
so a coupling constant renormalization is required. We now must
examine whether the propagator for $\sigma$ depends on this
renormalization.  We find that the two point function for $\sigma$ is
proportional to the inverse of
$$
 \frac{(p^2-4M^2)N}2\int\frac{d^Dk}{(2\pi)^D}
 \frac{1}{(k^2-M^2)[(k+p)^2-M^2]}\,\,\label{18},
$$
which is identical to the Gross-Neveu case. Notice that the gap equation
was not used. The finiteness of the above expression is a consequence
of supersymmetry. 

The interacting part of the noncommutative version of the
supersymmetric nonlinear sigma model 
is given by \cite{Girotti:2001gs}
\begin{eqnarray}
L_i &=& \frac{\lambda}{2}\star\varphi_i \star\varphi_i 
  - \frac12 F_i \star (\sigma\star\varphi_i +\varphi_i
\star\sigma) \nn
&-& \frac12 \sigma\star\overline\psi_i \star \psi_i  
-\frac12 (\bar\xi\star\psi_i \star\varphi_i
+\bar\xi\star\varphi_i\star\psi_i ) \nonumber\\
&-&\frac{N}{2g}\lambda - \frac{N M\sigma}{g}.
\label{19}
\end{eqnarray}
Notice that supersymmetry dictates the form of the trilinear
vertices. Also, the supersymmetry transformations are not modified by
noncommutativity since they are linear and no Moyal products are
required. 

The propagators are the same as in the commutative case. The vertices
have cosine factors due to the Moyal product.
We again consider the propagators for the Lagrange multiplier
fields. Now the $\sigma$ propagator is modified
by the cosine factors and is well behaved both in UV and IR regions.
The propagators for
$\lambda$ and $\xi$ are also well behaved in UV and IR regions.

The degree of superficial divergence for a generic 1PI graph $\gamma$ is
$
d(\gamma)= D - {(D-1)}N_\psi/2 - {(D-2)} N_\varphi/2 - {D}
N_F/2 - N_\sigma- 3 N_\xi/2 - 2 N_\lambda,\label{24}
$
where $N_{\cal O}$ is the number of external lines associated to the 
field ${\cal O}$. Potentially dangerous diagrams are those contributing
to the self--energies of the $\varphi$ and $\psi$ fields since, in principle,
they are quadratic and linearly divergent, respectively.
For the self-energies of $\varphi$ and $\psi$  we find that they
diverge logarithmically and they can be removed by a wave function
renormalization of the respective field. The same happens for the
auxiliary field $F$. The renormalization factors for them are the same
so supersymmetry is preserved in the noncommutative theory. 
This analysis can be extended to the n-point functions. In $2+1$
dimensions we find nothing new showing the renormalizability of the
model at leading order of $1/N$. However, in $1+1$ dimensions there
are 
some peculiarities. Since the scalar field is dimensionless in $1+1$
dimensions any graph involving an arbitrary number of external
$\varphi$ lines is quadratically divergent. In the four-point function
there is a partial cancellation of divergences but a logarithmic
divergence still survives. The counterterm needed to remove it can not
be written in terms of $\int d^2 x \,\,\varphi_i \star \varphi_i \star
\varphi_j\star\varphi_j$ and $\int d^2 x \,\,
\varphi_i\star\varphi_j\star\varphi_i\star\varphi_j$. A possible way 
to remove this divergence is by generalizing the definition of 1PI
diagram. However the cosine factors do not
allow us to use this mechanism which casts doubt about the
renormalizability of the noncommutative supersymmetric $O(N)$
nonlinear sigma model in $1+1$ dimensions. 

The noncommutative supersymmetric nonlinear sigma model can also be
formulated in superspace where it possible to show that model is
renormalizable to all orders of 1/N and explicitly 
verify that it is asymptotically free \cite{Girotti:2001ku}.   

\section{Spontaneous Symmetry Breaking in Noncommutative Field Theory}

Having seen the important role supersymmetry plays in noncommutative
models it is natural to go further. Spontaneous symmetry breaking and
the Goldstone theorem are essential in the 
standard model and the effect of noncommutativity in this setting
deserves to be fully 
understood. In four dimensions it is known that spontaneous symmetry
breaking can occur for the $U(N)$ model but not for the $O(N)$ unless
$N = 2$.  The $O(2)$ case was analyzed in detail
\cite{Campbell:2000ug} and 
the results for the $U(N)$ case have been extended to two loops
\cite{Liao:2002rp}. Going to higher loops requires an IR regulator
which can no longer be removed \cite{Sarkar:2002pb}. Due to these
troubles we will consider three dimensional models. 

Let us consider the three-dimensional Lagrangian\cite{Girotti:2002kr}  
\bea
\label{action1}
&L& = \frac{1}{2} \partial^\mu \phi_a\ \partial_\mu
\phi_a+\frac{\mu^2}{2}\phi_a\phi_a \nn  
&-&\frac{g}4  ( l_1 \phi_a*\phi_a*\phi_b*\phi_b
+l_2\phi_a*\phi_b*\phi_a*\phi_b) \nn
&-&
\frac{\lambda}{6} (h_1\phi_a*\phi_a*\phi_b*\phi_b*\phi_c*\phi_c+
\nn
&+&h_2\phi_a*\phi_a*\phi_b*\phi_c*\phi_c*\phi_b+ \nn
&+&
h_3\phi_a*\phi_a*\phi_b*\phi_c*\phi_b*\phi_c+ \nn
&+& h_4\phi_a*\phi_b*\phi_c*\phi_a*\phi_b*\phi_c+\nn
&+&
h_5\phi_a*\phi_b*\phi_c*\phi_a*\phi_c*\phi_b
) ,
\end{eqnarray}
where $l_1,l_2,h_1,h_2\ldots h_5$ are real numbers satisfying the conditions
$l_1+l_2=1$ and $h_1+h_2+\ldots+h_5=1$, so that there are two quartic
and five sextuple independent interaction couplings. The potential has
a minimum for $\phi_a\phi_a=a^2$ with 
$
a^2=  ( -g + \sqrt{g^2+ 4\mu^2\lambda} )/{2\lambda}.
$
As usual we introduce the field $\pi_i$ and $\sigma$ having vanishing
expectation value. 
Notice that the equation for $a^2$ implies that the pions 
are massless in the tree approximation, in agreement with the
Goldstone theorem. 


The gap equation receives no contribution from noncommutativity while
the one loop corrections to the pion mass have divergences both, in the
planar and non-planar sectors. Eliminating the UV divergence in the
planar sector also eliminates the UV/IR mixing in the non-planar
sector. It is also fortunate that it leads to an analytic behavior in
the IR so that the mass corrections vanish for $p=0$. This mechanism
does not appears in the four dimensional case. 

The two point function for $\sigma$ is also analytic in the IR leading
to a relation among the parameters. The divergences in the higher
point functions can also be eliminated. Therefore, we have shown that
this $O(N)$ model is renormalizable at one loop for any $N$
\cite{Girotti:2002kr}, in contradistinction to the four dimensional
case where $N$ must be equal to 2.  

A supersymmetric version of this model can be formulated in
superspace. Again, the gap equation is not affected by
noncommutativity. The mass corrections for the pion two point function
are UV finite and
free of UV/IR mixing as expected. It also vanishes for
$p=0$. Supersymmetry does not appear to be important in
this situation. 

\section{Noncommutativity and Gravity}

An important property of NC theories, which distinguishes them from
the conventional ones, is that translations in the NC
directions are equivalent to gauge transformations
\cite{Gross:2000ph}. This can be seen even for the case of a scalar
field which has the gauge transformation
$ \delta \hat{\phi} = -i [ \hat{\phi}, \hat{\lambda} ]_\star$,
where $[A,B]_\star  = A \star B - B \star A$ is the Moyal
commutator. Under a global
translation the scalar field transforms as $\delta_T \phi = \xi^\mu
\partial_\mu \hat{\phi}$. Derivatives of the field can be rewritten
using the Moyal commutator as $\partial_\mu \hat{\phi} = - i
\theta^{-1}_{\mu\nu} [ x^\nu, \hat{\phi}]_\star$ so that
$\delta \hat{\phi} = \delta_T \hat{\phi}$ with gauge parameter
$\hat{\lambda} = - \theta^{-1}_{\mu\nu} \xi ^\mu x^\nu$. The only other
field theory which has this same property is general relativity where
local translations are gauge transformations associated to general
coordinate transformations. This remarkable property shows that, as in
general relativity, there are no local gauge invariant observables in
NC theories.

As remarked in the introduction the connection between
translations and gauge transformations seems to be lost after the SW
map. A global 
translation on commutative fields can no longer be rewritten as a gauge
transformation. We will show that another aspect
concerning gravity emerges when commutative fields are
employed. Noncommutative field theories can be interpreted as ordinary
theories immersed in a gravitational background generated by the gauge
field \cite{Rivelles:2002ez}.

The action for a real scalar field in the adjoint representation of
$U(1)$ coupled minimally to a gauge field and in flat space-time is 
$$
S_\varphi = \frac{1}{2} \int d^4x \,\,\, \hat{D}^\mu \hat{\varphi} \star
\hat{D}_\mu \hat{\varphi},
$$
where $\hat{D}_\mu \hat{\varphi} = \partial_\mu \hat{\varphi} - i
[\hat{A}_\mu, \hat{\varphi} ]_\star$. The SW map is given by 
$
\hat{A}_\mu = A_\mu - \frac{1}{2} \theta^{\alpha\beta}
A_\alpha ( \partial_\beta A_\mu + F_{\beta\mu} ), \nn
\hat{\varphi} = \varphi - \theta^{\alpha\beta} A_\alpha
\partial_\beta \varphi, 
$ so that the action can be written, to first order in $\theta$, as
\bea
&&S_\varphi = \frac{1}{2} \int d^4x \, [ \partial^\mu \varphi \partial_\mu
\varphi + \nn
&& 2 \theta^{\mu\alpha} {F_\alpha}^\nu ( - \partial_\mu \varphi
\partial_\nu \varphi + \frac{1}{4} \eta_{\mu\nu} \partial^\rho \varphi
\partial_\rho \varphi ) ]. 
\eea
Notice that the tensor inside the parenthesis is
traceless. If we now consider this same field coupled to a
gravitational background and expand the metric around the flat
Minkowski metric $\eta_{\mu\nu}$ as $g_{\mu\nu} = \eta_{\mu\nu} +
h_{\mu\nu} + \eta_{\mu\nu} h,$ where $h_{\mu\nu}$ is traceless, we get
\be
\label{linearized_scalar_action}
L = \frac{1}{2} ( \partial^\mu \varphi
\partial_\mu \varphi - h^{\mu\nu}  \partial_\mu \varphi \partial_\nu
\varphi + 
 h \partial^\rho \varphi \partial_\rho \varphi ),
\ee
where indices are raised and lowered with the flat metric.
Since both actions have the same structure we can
identify a linearized background gravitational field
$
h^{\mu\nu} =  \theta^{\mu\alpha} {F_\alpha}^\nu + \theta^{\nu \alpha}
{F_\alpha}^\mu + \frac{1}{2} \eta^{\mu\nu} \theta^{\alpha\beta}
F_{\alpha\beta}$ and 
$h = 0.
$
This linearized metric describes a gravitational plane wave
\cite{Rivelles:2002ez}. Then, the effect of noncommutativity on the
commutative scalar field 
is similar to a field dependent gravitational field.

The same procedure can be repeated for the complex scalar field and we
get a linearized contribution that is half of that of the real scalar
field. Then charged fields feel a gravitational background which is half of
that felt by the uncharged ones. Therefore, the gravity coupling is now
dependent on the charge of the field, being stronger for uncharged
fields. Notice that the gauge field has now a dual role, it couples
minimally to the charged field and also as a gravitational
background. 

We can now turn our attention to the behavior of a charged massless
particle in this background. Its geodesics is described by
$$
\label{ds}
( 1 + \frac{1}{4} \theta^{\alpha\beta} F_{\alpha\beta} ) dx^\mu dx_\mu
+ \theta_{\mu\alpha} {F^\alpha}_\nu dx^\mu 
dx^\nu = 0.
$$
If we consider the case where there is no noncommutativity between
space and time, that is $\theta^{0i}=0$, and
calling $\theta^{ij} = \epsilon^{ijk} \theta^k$, $F^{i0} = E^i$, and
$F^{ij} = \epsilon^{ijk} B^k$, we find to first order in $\theta$ that
$
( 1 -  \vec{v}^2 )( 1 - 2 \vec{\theta} \cdot \vec{B}) - \vec{\theta}
\cdot (\vec{v} \times \vec{E} ) + \vec{v}^2 \vec{\theta} \cdot \vec{B}
- (\vec{B} \cdot \vec{v}) ( \vec{\theta} \cdot \vec{v}) = 0,
$
where $\vec{v}$ is the particle velocity. Then to zeroth order, the
velocity $\vec{v}_0$ satisfies
$\vec{v}_0^2 = 1$ as it should. We can now decompose all vectors into
their transversal and longitudinal components with respect to
$\vec{v}_0$, $\vec{E} = \vec{E}_T + \vec{v}_0 E_L$, $\vec{B} =
\vec{B}_T + \vec{v}_0 B_L$ and $\vec{\theta} = \vec{\theta}_T + \vec{v}_0
\theta_L$. We then find that the velocity is
\be
\label{velocity_real_scalar}
\vec{v}^2 = 1 + \vec{\theta}_T \cdot ( \vec{B}_T - \vec{v}_0 \times
\vec{E}_T ).
\ee
Hence, a charged massless particle has its velocity changed with
respect to the velocity of light by an amount which depends on
$\theta$. 

We can now check the consistency of these results by going back to the
original actions and 
computing the group velocity for planes waves. Upon quantization
they give the velocity of the particle associated to the
respective field. For the charged scalar field we get the equation
of motion
$( 1 - \frac{1}{4} \theta^{\mu\nu} F_{\mu\nu}) \partial^2
\varphi - \theta^{\mu\alpha} {F_\alpha}^\nu \partial_\mu\partial_\nu
\varphi = 0.
$
If the field strength is constant we can find a plane wave solution
with the following dispersion relation
$
\frac{\vec{k}^2}{\omega^2} = 1 -  \vec{\theta}_T \cdot ( \vec{B}_T -
\frac{\vec{k}}{\omega} \times \vec{E}_T ),
$
where $k^\mu = (\omega, \vec{k})$.
We then find that the phase and group velocities coincide and are given
by (\ref{velocity_real_scalar}) as expected. Therefore, in both pictures,
noncommutative and gravitational, we get the same results. 
The dispersion relation here is similar to that found for
photons in a background 
magnetic field \cite{Guralnik:2001ax}. 

This work was partially supported by CAPES, CNPq and PRONEX
under contract CNPq 66.2002/1998-99.

\end{document}